\documentclass[10pt,journal,doublecolumn]{IEEEtran}

\usepackage{threeparttable,placeins,makecell,stfloats,cite}
\usepackage{amsmath,amssymb,cite}
\usepackage[dvips]{graphicx}
\usepackage[usenames]{color}
\usepackage{amsfonts}
\usepackage{latexsym}
\usepackage{multirow}
\usepackage{subfigure}
\usepackage{float}

\usepackage[format=hang]{subfig}
\usepackage{threeparttable,placeins,makecell,stfloats,cite}

\ifCLASSINFOpdf
 \else
 \fi

\begin{document}
\title{Power-Imbalanced Low-Density Signatures (LDS) From Eisenstein Numbers}
\author{\IEEEauthorblockN{Zilong Liu, Pei Xiao, and Zeina~Mheich}
\IEEEauthorblockA{\\Institute of Communication Systems,\\ Home of 5G Innovation Centre,\\ University of Surrey, UK.\\
Email: {\tt \{zilong.liu,p.xiao,z.mheich\}@surrey.ac.uk}}}
 \maketitle

 \begin{abstract}
As a special case of sparse code multiple access (SCMA), low-density signatures based code-division multiple access (LDS-CDMA) was widely believed to have worse error rate performance compared to SCMA. With the aid of Eisenstein numbers, we present a novel class of LDS which can achieve error rate performances comparable to that of SCMA in Rayleigh fading channels and better performances in Gaussian channels. This is achieved by designing power-imbalanced LDS such that variation of user powers can be seen both in every chip window and the entire sequence window. As LDS-CDMA is more flexible in terms of its backwards compatibility, our proposed LDS are a promising sequence candidate for dynamic machine-type networks serving a wide range of communication devices.
\end{abstract}

\vspace{0.1in}

\section{Introduction}
The fifth-generation communication systems (5G) and beyond need to deal with explosive growth of communication devices which are connected to provide seamless and ubiquitous wireless data services. These devices, e.g., smart meters in factories of future (FoF), may be densely deployed in certain area for a diverse range of data collection and/or control operations. The devices generally remain in sleep mode with the exception of short periods of time during which small data packets are exchanged in a sporadic way. The communications over such massive number of communication devices are called machine-type communications (MTC). This paper is focused on the multiple access design for downlink MTC systems to support reliable and flexible massive connectivity. 

LDS-CDMA is a code-domain non-orthogonal multiple access (NOMA) technique in which the multiuser detection (MUD) is conducted by efficiently exploiting the sparsity of LDS with message passing algorithm (MPA) \cite{LDS2008}. In an LDS-CDMA system, each user spreads its data symbols by a unique LDS whose sequence entries are mostly zero. The concept of LDS-CDMA was soon extended to SCMA, in which each user sends out a sparse codeword (from a properly designed sparse codebook) based on the instantaneous input message \cite{SCMA2013,SCMA2014}. The error rate performance of SCMA benefits from the so-called ``constellation shaping gain" (owned by its sparse codebooks). Hence, it was widely believed in the literature that LDS-CDMA gives rise to worse error rate performance compared to SCMA.

Numerous constructions have been proposed for LDS and SCMA codebooks \cite{Beek2009,Yu2015,Cai2016,Bao2016,Yan2016,Zhou2017,Song2017,Wei2017,Yu2018,Zeina2019}. Most existing works on SCMA codebook design start from a single multi-dimensional mother constellation having large minimum product distance (or minimum Euclidean distance) \cite{Boutros1996,Boutros1998}, with which sparse codebooks are generated for multiple users through constellation operations such as phase rotations, interleaving, and permutations. These operations lead to power-imbalanced constellations, i.e., variation of user powers can be seen from sparse codebooks pertinent to each resource node. Power-imbalanced constellations help to amplify the ``near-far effect" which is useful for enhanced interference cancellation in MPA. In contrast, existing LDS constructions are generally focused on ``power-balanced" LDS \cite{Beek2009,Zhou2017,Song2017}, i.e., all the non-zero sequence entries connected to every resource node possess identical magnitude. So far, the design of \textit{optimal} LDS and SCMA codebooks is still an open problem.

In this work, we propose a novel class of power-imbalanced LDS with the aid of Eisenstein numbers. Our proposed LDS have the property that power variation can be seen 1) at every chip\footnote{``Chip" is a concept in CDMA theory. Each LDS chip, sent over a time- or frequency- slot, refers to a sequence entry.} window (i.e., all the active users in a chip window are given different power levels) and 2) at the entire sequence window (i.e., all the users may be allocated with different powers). By doing so, we show that larger minimum product distance can be attained by the proposed LDS. Simulations indicate that our proposed LDS can achieve error rate performances comparable to the Huawei SCMA codebook in \cite{HuaweiCodebook} in Rayleigh fading channels and better performances in Gaussian channels.

\section{Introduction to SCMA}

 \begin{figure*}[htbp]
  \centering
  \includegraphics[width=5in]{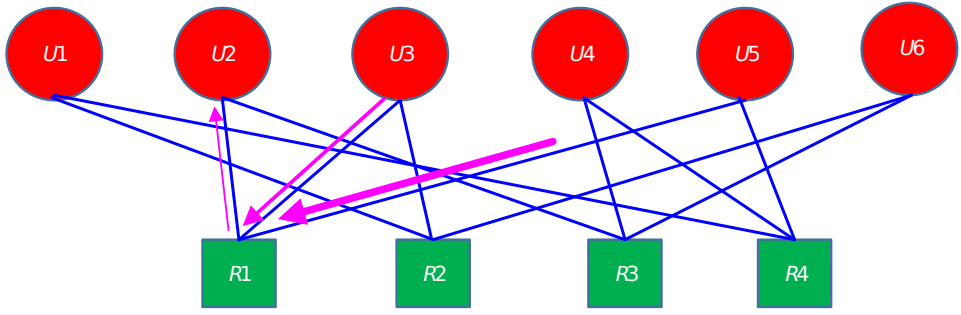}\\
  \caption{Factor graph of an SCMA system with $J=6,K=4,d_v=2,d_c=3$.}
  \label{factor-graph}
\end{figure*}
We consider a downlink SCMA system in which the basestation simultaneously communicates with $J$ users sharing $K$ orthogonal resources. In practice, the SCMA signals are transmitted over an orthogonal frequency-division multiplexing (OFDM) system with $K$ subcarriers. Each user is given a multi-dimensional sparse codebook consisting of $M$ codewords with identical length of $K$. Such a codebook may be written as a $K\times M$ matrix, denoted by $\mathbf{X}_j,j\in \{1,2,\cdots,J\}$. The SCMA encoder for user $j$ selects a codeword, denoted by $\mathbf{x}_j$ which is essentially a column of $\mathbf{X}_j$, based on the instantaneous input message consisting of $\log_2(M)$ bits. Sparse codebooks of an SCMA system can be characterised by a bipartite factor graph consisting of resource nodes and user nodes. We consider a regular factor graph where each user node has degree of $d_v$ and each resource node has degree of $d_c$. Due to the sparsity, each codebook $\mathbf{X}_j$ is comprised of $K-d_v$ zero rows. In SCMA, in general, we have $J>K$, meaning that the number of users that can be simultaneously transmitted is larger than the total number of orthogonal resources. For ease of presentation, we define $J/K$  as the overloading factor of an SCMA system.

Fig. \ref{factor-graph} illustrates the factor graph of an SCMA codebook with $J=6,K=4,d_v=2,d_c=3$. In Fig. \ref{factor-graph}, each circle (in red) represents a user node, while each square box (in green) represents a resource node. The arrows (in purple) in Fig. \ref{factor-graph} show how the soft messages are passed during the MPA at the receiver. An alternative method of representing the factor graph is by an indicator matrix, in which each row denotes a specific resource node and all the non-zero entries in the row give the users which have active transmissions over this resource node. Following this principle, the above factor graph can be represented by the indicator matrix as follows:
\begin{equation}\label{ind_matrix_equ}
\mathbf{F}=\left [
\begin{matrix}
0 & 1 & 1 & 0 & 1 & 0\\
1 & 0 & 1 & 0 & 0 & 1\\
0 & 1 & 0 & 1 & 0 & 1\\
1 & 0 & 0 & 1 & 1 & 0
\end{matrix}
\right ].
\end{equation}	

In LDS-CDMA, as a special case of SCMA, one may also characterize it by sparse codebooks each of which can be expressed as the Kronecker product of a length-$K$ sparse sequence (with $d_v$ nonzero sequence entries due to the sparsity), denoted by $\mathbf{s}_j$, and a constellation set of order $M$. Specifically,
\begin{equation}
\mathbf{X}_j=[\mathbf{s}_j\alpha_1,\mathbf{s}_j \alpha_2,\cdots,\mathbf{s}_j \alpha_M],
\end{equation}
where $\{\alpha_1,\alpha_2,\cdots,\alpha_M\}$ refers to the constellation set. Hence, the rank of $\mathbf{X}_j$ in LDS is equal to one only.

In the downlink transmissions, the received signal $\mathbf{y}$ can be expressed as
\begin{equation}
\mathbf{y}=\sum\limits_{j=1}^J \text{diag}(\mathbf{h})\mathbf{x}_j+\mathbf{n},
\end{equation}
where $\mathbf{h}=[h_1,h_2,\cdots,h_K]^T$ denotes the channel fading vector with $h_k\sim\mathcal{CN}(0,1)$ and $\mathbf{n}=[n_1,n_2,\cdots,n_K]^T$ denotes the additive white Gaussian noise (AWGN) vector with $n_k\sim\mathcal{CN}(0,N_0)$.
Let
\begin{displaymath}
\mathbf{x}=\sum_{j=1}^J \mathbf{x}_j
\end{displaymath}
be the superimposed codeword over each SCMA transmission. In total, there are $M^J$ superimposed codewords, which form a set denoted by $\mathcal{X}$, when all the possible input messages from all the $J$ users are enumerated. In Rayleigh fading channels, the error rate performance of an SCMA system is determined by the pairwise error probability (PEP) which is upper bounded by \cite{Boutros1996,Boutros1998}
\begin{equation}
\text{Prob}(\mathbf{x}\rightarrow \mathbf{y})\leq \frac{1}{2}\prod_{x_k\neq y_k}\frac{1}{1+\frac{|x_k-y_k|^2}{8N_0}},
\end{equation}
where $\mathbf{x},\mathbf{y}$ denote the transmitted superimposed codeword and the superimposed codeword decoded by the receiver, respectively.
 By viewing the superimposed codeword set $\mathcal{X}$ as a multidimensional constellation, SCMA codeooks and LDS in Rayleigh fading channels can be optimized by the following ways:
\begin{enumerate}
\item Maximizing the diversity order $L$ of $\mathcal{X}$;
\item Maximizing the minimum product distance square (MPDS) defined below:
\begin{displaymath}
d_p=\min_{\mathbf{x},\mathbf{y}\in \mathcal{X}} \prod_{x_k\neq y_k} |x_k-y_k|^2;
\end{displaymath}
\item Minimizing the kissing number of the $\mathcal{X}$.
\end{enumerate}
It is noted that the diversity order of SCMA system is upper bounded by $d_v$ \cite{Lim2017} which can be easily met by design. The optimization of the kissing number, however, is not so straightforward. Therefore, maximizing the MPDS may be a feasible way for enhanced SCMA codebooks and LDS.

\section{Proposed LDS From Eisenstein Number}

\subsection{Proposed LDS Construction}
\begin{figure}[htbp]
  \centering
  \includegraphics[width=3.2in]{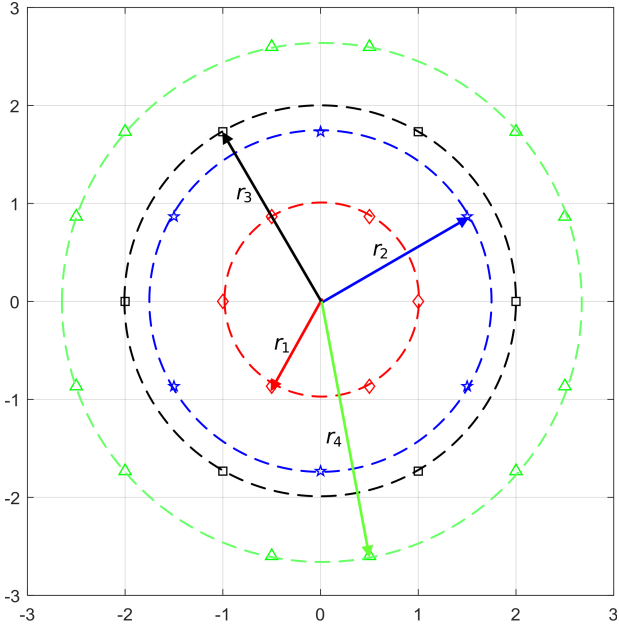}\\
  \caption{Eisenstein integers over the first four rings, where $r_1=1,r_2=\sqrt{3},r_3=2,r_4=\sqrt{7}$.}
  \label{LDS-consti}
\end{figure}

Let us consider the Eisenstein number below.
\begin{equation}
\omega=\exp\left (\frac{2\pi i}{3}\right ),
\end{equation}
where $i=\sqrt{-1}$.
The above Eisenstein number is a root of $x^2+x+1=0$. As a matter of fact, $\{a+b\omega:a,b \in \mathbb{Z}\}$, called Eisenstein integers, form the hexagonal lattice in the complex domain. In the theory of data packing, a hexagonal lattice is viewed as the best lattice for its largest packing distance \cite{Proakis2008}. Fig. \ref{LDS-consti} portraits the Eisenstein numbers over the first four rings, where the ring radiuses are $r_1=1,r_2=\sqrt{3},r_3=2,r_4=\sqrt{7}$, respectively.

Let us define $\zeta_k,1\leq k\leq K$, which is a set containing the indices of users which are active at resource node $k$. For an SCMA system with regular factor graph, the cardinality of $\zeta_k$ is $d_c$. For the SCMA system characterized by the sparse indicator matrix in (\ref{ind_matrix_equ}),  we have $\zeta_1=\{2,3,5\},\zeta_2=\{1,3,6\},\zeta_3=\{2,4,6\},\zeta_4=\{1,4,5\}$.

Denote by $\mathbf{S}$ a $K\times J$ sparse matrix formed by LDS, where
\begin{equation}
\begin{split}
\mathbf{S} & =\left [
\begin{matrix}
\mathbf{s}_{1},&\mathbf{s}_2,&\cdots,&\mathbf{s}_J
\end{matrix}
\right ],\\
\mathbf{s}_j &= [s_{j,1},s_{j,2},\cdots,s_{j,K}]^T,1\leq j\leq J.
\end{split}
\end{equation}
Due to the sparsity, we have $s_{j,k}=0$ if the $(j,k)$-element of the indicator matrix $\mathbf{F}$ is zero, i.e., $F_{j,k}=0$. Hence, the positions of all the nonzero entries of the $k$-th row of $\mathbf{S}$ are given by $\zeta_k$.

\vspace{0.1in}

\noindent \textit{Proposed LDS Construction}:

\noindent \textit{Step 1}: Based on the hexagonal lattice formed by Eisenstein numbers, select $d_c$ rings whose radiuses satisfying
\begin{displaymath}
0<R_1<R_2<\cdots<R_{d_c}.
\end{displaymath}
The $t$-th ring consists of all the Eisenstein numbers with magnitude of $r_t$, where $1\leq t \leq d_c$. For ease of presentation, denote the Eisenstein numbers lying on the $t$-th ring by set $\xi_t$.

\noindent \textit{Step 2}: Find a sparse matrix $\mathbf{S}$ such that every row contains $d_c$ nonzero entries which are Eisenstein numbers with different magnitudes lying on the $d_c$ rings obtained in \textit{Step 1}, i.e., each nonzero entry is drawn from a distinctive set $\xi_t(1\leq t \leq d_c)$.

\noindent \textit{Step 3}: Normalize the sparse matrix obtained in \textit{Step 2} such that $\parallel \mathbf{S} \parallel^2_F=d_vJ$.

\vspace{0.1in}
It is noted that the power imbalance of among chips of LDS is introduced in the above \textit{Step 2}. As all the $d_c$ nonzero entries associated to each row of $\mathbf{S}$ have different magnitudes,
the $d_c$ active users over each resource node are assigned with different power levels. As will be shown by the LDS example in the next subsection, the power imbalance can also be introduced among different users. It is also pointed out that existing works on LDS design generally assumes uniform power allocation among all the users, i.e., $\parallel \mathbf{s}_j \parallel^2_F=d_v$ for all $1\leq j\leq J$. However, by discarding this constraint in the above \textit{Step 2}, we will show next that LDS with large MPDS is possible.
\vspace{0.1in}

\subsection{Example and Numerical Analysis}
Based on the factor graph in Fig. \ref{factor-graph} and with the aid of computer, we obtain the following set of LDS with large MPDSs in (\ref{LDS_equ}). Specifically,
$\mathbf{S}_1$ is constructed by setting $R_1=r_1,R_2=r_2,R_3=r_4$. The energy distributions of $\mathbf{S}_1$ are
\begin{displaymath}
[ 2.1818,~1.6364,~1.0909,~2.1818,~2.7273,~2.1818].
\end{displaymath}
Hence, power imbalance can be seen not only at each resource node (i.e., chip-level), but also among different users.
Setting the 4-point Huawei codebook in \cite{HuaweiCodebook} with MPDS of $0.0005$ as a reference, we observe that under QPSK spreading, the MPDS of $\mathbf{S}_1$ is $0.0144$. Moreover, we consider the set of conventional LDS with MPDS of $0.0091$, denoted by $\mathbf{S}_2$ in (\ref{LDS_equ}), which is generated according to the ``Latin" rule in \cite{Beek2009}.

\begin{figure*}
\begin{equation}\label{LDS_equ}
\begin{split}
\mathbf{S}_1 & =\left [
\begin{array}{cccccc}
   {\scriptstyle 0}                  &  {\scriptstyle 0.7833 - 0.4523i}            &  {\scriptstyle -0.5222} &{\scriptstyle  0                 }& {\scriptstyle 1.3056 - 0.4523i }&{\scriptstyle  0}\\
   {\scriptstyle -0.2611 - 1.3568i}   &  {\scriptstyle 0}                 &  {\scriptstyle -0.7833 + 0.4523i} &{\scriptstyle  0                 }& {\scriptstyle 0                }&{\scriptstyle -0.2611 - 0.4523i}\\
   {\scriptstyle 0}                  &  {\scriptstyle 0.9045i} &  {\scriptstyle 0               } &{\scriptstyle 0.2611 + 0.4523i}& {\scriptstyle 0                }&{\scriptstyle  0.2611 + 1.3568i}\\
   {\scriptstyle -0.5222}   &  {\scriptstyle 0}                 &  {\scriptstyle 0               } &{\scriptstyle 1.0445 + 0.9045i  }& {\scriptstyle -0.7833 - 0.4523i           }&{\scriptstyle  0}
\end{array}
\right ],\\
\mathbf{S}_2 & = \left [
\begin{array}{cccccc}
 0            & \omega      & 1      & 0           & \omega^{1/2} & 0\\
 \omega^{1/2} & 0           & \omega & 0           & 0            & 1\\
 0            & \omega^{1/2}& 0      & 1           & 0            & \omega\\
 1            & 0           & 0      & \omega^{1/2}& \omega       & 0
\end{array}
\right ].
\end{split}
\end{equation}
\end{figure*}

\vspace{0.1in}

%

To analyze the error rate performance, we evaluate the average bit error rate (ABER) of the two sets of LDS (i..e, $\mathbf{S}_1,\mathbf{S}_2$) and the Huawei codebook. The average BER (denoted by $P_{\text{avg}}$) is evaluated by counting the error rates of all the possible superimposed codwords. It can be readily shown that
\begin{equation}
\begin{split}
P_{\text{avg}} & \leq \frac{1}{M^J}\sum_{\mathbf{x}\in \mathcal{X}}\sum_{\mathbf{y}\neq \mathbf{x},\mathbf{y}\in \mathcal{X}}\frac{d_H(\mathbf{x},\mathbf{y})}{J\log_2(M)}\text{Prob}(\mathbf{x}\rightarrow \mathbf{y}),
\end{split}
\end{equation}
where $d_H(\mathbf{x},\mathbf{y})$ denotes the number of different information bits associated to superimposed codewords $\mathbf{x}$ and $\mathbf{y}$.

\begin{figure}[htbp]
  \centering
  \includegraphics[width=3.6in]{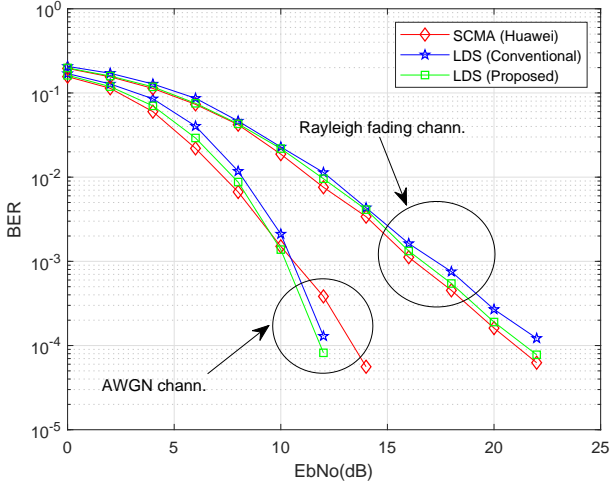}\\
  \caption{ABER comparison for the two LDS sets (with QPSK spreading) and the Huawei codebook (4-point SCMA).}
  \label{BERComp}
\end{figure}

Fig. \ref{BERComp} shows the ABER comparison over Rayleigh fading channels and AWGN channels. It is interesting to see that our proposed power-imbalanced LDS set $\mathbf{S}_1$ gives rise to ABER very close to that of the SCMA codebook in Rayleigh fading channels. In contrast, the power-balanced set $\mathbf{S}_2$ gives rise to ABER which is about 1 dB away from that of the SCMA codebook for ABER at $10^{-4}$. In AWGN channels, both LDS sets outperform the SCMA codebook in high EbNo region, with the largest performance gain achieved by the proposed power-imbalanced LDS set.

\section{Conclusions}
In this paper, we have proposed a new class of LDS which can achieve error rate performances comparable to that of SCMA in Rayleigh fading channels and better error rate performances in AWGN channels. Our main idea is to first generate a set of Eisenstein numbers lying over rings with different radiuses and then carefully select nonzero sequence entries from these rings such that the constructed LDS exhibit power imbalance (and larger MPDS) in all the chip windows and the entire sequence window. It should be pointed out that LDS design similar to the proposed one may be carried out based on other rings such as amplitude-phase-shift-keying (APSK) constellations. A deeper investigation on optimal selection of the constellation rings will be a future work of this research.

We remark here that LDS-CDMA is more flexible than SCMA in that the decoding of LDS-CDMA may also be carried out by MUD techniques in conventional CDMA theory (e.g., matched filters based MUD). Two typical scenarios where conventional MUD techniques may be useful are:
\begin{enumerate}
\item The MPA decoding in SCMA receiver generally assumes that all the $J$ active users communicate over $K$ resource nodes. In an MTC network with dynamic traffic flow, the MPA decoding performance may not be guaranteed when the number of users reduces to less than the maximum number $J$.
\item For an MTC network with a diverse range of communication devices in terms of their computation resources, the MPA decoding may be too expensive to afford as the iterative decoding tends to consume more receiver power.
\end{enumerate}
In short, our proposed LDS provide a promising sequence candidate to enhance the flexibility and backwards compatibility of MTC networks.
%
%

\end{document}